\documentclass{article}
\usepackage{graphics}

\begin{document}

\title{Finding Routes in Anonymous Sensor Networks}

\author{Renato~C.~Dutra\\
Valmir~C.~Barbosa\thanks{Corresponding author
({\tt valmir@cos.ufrj.br}).}\\
\\
Universidade Federal do Rio de Janeiro\\
Programa de Engenharia de Sistemas e Computa\c c\~ao, COPPE\\
Caixa Postal 68511\\
21941-972 Rio de Janeiro - RJ, Brazil}

\date{}

\maketitle

\begin{abstract}
We consider networks of anonymous sensors and address the problem of
constructing routes for the delivery of information from a group of sensors in
response to a query by a sink. In order to circumvent the restrictions imposed
by anonymity, we rely on using the power level perceived by the sensors in the
query from the sink. We introduce a simple distributed algorithm to achieve the
building of routes to the sink and evaluate its performance by means of
simulations.

\end{abstract}

\section{Introduction}\label{intro}

A sensor network is a wireless network of simple elements, called sensors, that
have sensing or monitoring capabilities related to some application domain, and
have in addition limited processing and communication capabilities. Sensors are
typically distributed irregularly in space and rely for operation on autonomous
power sources that in general cannot be recharged, so expending energy as
minimally as possible is a crucial concern. Sensor networks are currently being
considered for use in a variety of contexts, ranging from biomedical to
environmental monitoring tasks, often involving otherwise inaccessible
monitoring sites or conditions that are too hazardous for direct human
involvement.

Several typical tasks that sensor networks are planned to perform involve the
use of more powerful processing and communicating elements, called sinks, that
also communicate by wireless means with the sensors but have in addition the
capability to connect to some outside network, like the Internet. In general,
sinks are not constrained to using power as economically as the sensors. One
common role performed by a sink is to broadcast a monitoring-related question to
a group of sensors and to relay a compilation of their replies through the
outside network for analysis. Because of the sensors' irregular spatial
distribution and limited power resources, conveying such replies to the sink
requires that clever routing and aggregation mechanisms be devised and has
inspired the development of techniques and algorithms at various protocol
levels. For recent reviews on such developments, we refer the reader to
\cite{assc02,ces04}.

Distributed algorithms for sensor-network operation invariably rely on the
assumption that each sensor can be, for all relevant purposes, uniquely
identified. This is a reasonable assumption to make: not only is it
technologically feasible \cite{hhkk04}, but also it has been known since the
seminal contributions in \cite{asw88} that there exist severe inherent
limitations to computing distributedly when the underlying processing elements
are anonymous. Such limitations have been established under the assumption that
point-to-point communication is available, and are therefore expected to be no
less stringent in the wireless scenario.

However, approaching distributed computing from the perspective of anonymity is
not so much a technology-related issue, but is rather a means of posing
questions aimed at highlighting a system's fundamental capabilities and
limitations. So the whole issue of distributed computing by anonymous elements
makes sense in the sensor-network context as well, even though it appears to
have remained untouched so far. In this paper we make what we think is the first
contribution to understanding how the assumption of anonymity impacts the
functioning of a sensor network.

We proceed in the following manner. In Section~\ref{algo} we introduce the
specific problem we address and give a simple distributed algorithm that
attempts to solve it. The core premise behind the design of this algorithm is
that the power level perceived by the sensors as they receive a transmission
from a sink can be used to provide some level of differentiation among them and
therefore compensate, to some degree, the assumption of anonymity. We then move
to a performance evaluation of the distributed algorithm in
Section~\ref{results} and finish in Section~\ref{concl} with conclusions.

\section{The problem and an algorithm}\label{algo}

We consider $n$ sensors placed arbitrarily in two-dimensional space and assume
the existence of one single sink. Sensors are assumed to have no
identifications, not even their coordinates in space. We assume that the sink
broadcasts one single question to all sensors and that $n^\star$ of the $n$
sensors are the ones to answer. We call each of these $n^\star$ sensors a source
and assume that sources are distributed arbitrarily amid the $n$ sensors. The
problem we address is the problem of finding routes from all sources to the
sink. Because sources can only broadcast at low power, their answers are likely
not to reach the sink directly but need instead to be routed through the other
sensors. All $n$ sensors, even though $n-n^\star$ of them do not have an answer
for the sink, may have a part to play in aggregating and relaying the sources'
answers.

The question the sink broadcasts reaches each sensor at a power level that is
inversely proportional to the square of its distance to the sink. The key
premise underlying our approach is that each sensor is capable of measuring the
amount of power it perceives in the sink's transmission. For sensor $i$, we
denote this measure by $P_i$. Clearly, if for sensors $i$ and $j$ we have
$P_i>P_j$, then $i$ is closer to the sink than $j$ is, provided that the sink's
broadcast reaches all sensors isotropically (that is, at the same power level
for the same distance from the sink), as we henceforth assume. While this is
obviously no means of telling sensors apart from one another, since it only
differentiates sensors radially with respect to the sink, we demonstrate in the
remainder of the paper that it is possible to use such a property to provide
routing from all sources to the sink.

In order to do this, we first introduce a simple distributed algorithm for
execution by the sink and the sensors. We give the algorithm in a parametric
form and in Section~\ref{results} provide simulation results that aim at
clarifying which parameter ranges and values provide the desired results. We
assume that an upper bound $R$ on the greatest distance from the sink to a
sensor is known to the sink. If $B_0$ is the power level at which the sink
broadcasts its question, then sensor $i$, upon being reached by this broadcast
and measuring $P_i$, can calculate its distance to the sink, denoted by $R_i$,
and also its radial distance to the circle of radius $R$ centered on the sink
(that is, $R-R_i$): all it takes is that the sink broadcast, along with its
question, the values of $B_0$ and $R$ \cite{lc70}. We let $T$ and $T_i$ denote
the propagation times of an electromagnetic wave over the distances $R$ and
$R-R_i$, respectively (these can be computed easily given the wave's speed in
the medium under consideration).

We describe our distributed algorithm loosely after the general template of
reactive actions normally used for asynchronous distributed algorithms
\cite{b96}. All we must specify is then the initial broadcast by the sink, the
action to be taken by a sensor upon receiving this message, and also how the
sink or a sensor reacts to receiving a message from a sensor. In our description
of the algorithm, we use $\mathcal{S}_0$ and $\mathcal{S}_i$ to denote the
(otherwise unspecified) data structure used respectively by the sink and sensor
$i$ to aggregate all information it receives. If sensor $i$ is a source, then
initially $\mathcal{S}_i$ is assumed to contain its answer to the sink's
question.

The description that follows is given in terms of Actions~1 and~2, respectively
for the sink and for a generic sensor $i$. Action~2, in particular, is
dependent upon the product $fr$ of the two parameters $f$ and $r$. These are,
respectively, a number in the interval $[0,1]$ and the radius that a broadcast
by a sensor is desired to reach. Once the value of $r$ is known, we assume that
sensors broadcast at a power level, the same for all sensors, such that the
locations at which the message can be received are exactly those that are no
farther apart from the sensor than $r$. We return to how the value of $r$ is
chosen in Section~\ref{results}.

\begin{description}
\item[Action 1.] The sink broadcasts $\mathit{Question}(B_0,R)$ and sets a timer
to go off $2T$ time units later. In the meantime, upon receiving a message
$\mathit{Answer}(*,\mathcal{S})$ the sink incorporates $\mathcal{S}$ into
$\mathcal{S}_0$. When the timer goes off, the sources' answers to the sink are
all summarized in $\mathcal{S}_0$.
\item[Action 2.] Upon receiving the message $\mathit{Question}(B_0,R)$, sensor
$i$ broadcasts $\mathit{Answer}(P_i,\mathcal{S}_i)$ if it is a source, and
regardless of being a source or not sets a timer to go off $2T_i$ time units
later. In the meantime, upon receiving a message
$\mathit{Answer}(P,\mathcal{S})$ sensor $i$ incorporates $\mathcal{S}$, suitably
tagged with $P$, into $\mathcal{S}_i$. When the timer goes off, sensor $i$
checks whether $\mathcal{S}_i$ has had any information incorporated into it from
an $\mathit{Answer}$ message. In the affirmative case, it selects from
$\mathcal{S}_i$ the entry whose $P$ tag is greatest among all entries that have
a $P$ tag such that $P<P_i$. If the selection is successful (i.e., there is at
least one candidate entry), then let $P_j$ be this greatest $P$ tag; sensor $i$
then calculates $R_j$ from $P_j$. If it is unsuccessful, then sensor $i$ lets
$R_j=\infty$. It then broadcasts $\mathit{Answer}(P_i,\mathcal{S}_i)$ if
$R_j-R_i>fr$.
\end{description}

For simplicity's sake, we have given these two actions under the further
assumption that local computation, channel acquisition, and message transmission
by sensors take only negligible time if compared to the time for wave
propagation given the distances involved in the application at hand. This is
reflected in the values timers are set to, but these can clearly be increased to
satisfaction if the assumption does not hold. What is intended with Actions~1
and~2 is then the following. All sources broadcast their answers upon being
reached by the sink's question. All sensors, source or otherwise, upon this same
event, set timers proportionally to the round-trip time to the circle of radius
$R$ centered on the sink. As the timers go off in succession from the circle's
outskirts inward, the sensors aggregate the answers they receive from farther
out and pass the result on toward the sink. It all culminates with the sink's
timer going off, at which time all activity has ceased and the sink has
collected a set of aggregated answers, hopefully including answers from all
sources.

Notice, in Action~2, that considering $\mathcal{S}_i$ entries whose $P$ tags are
such that $P<P_i$ excludes data received from sensors that are nearer the sink
than sensor $i$. These excluded sensors are necessarily sources, since these are
the only sensors that broadcast their answers independently of timers.
Proceeding in this way is meant to prevent the progressive convergence of
aggregated information onto the sink from being interrupted: by Action~2, a
source's answer may go into $\mathcal{S}_i$ for some sensor $i$ farther away
from the sink, and proceeding differently would cause this sensor not to
participate in the process, i.e., not to send an $\mathit{Answer}$ message when
its timer went off (the $P$ tag of that source's data in $\mathcal{S}_i$ would
be such that $P>P_i$ and thus lead to $R_j<R_i)$.

\begin{figure}
\centering
\scalebox{1.000}{\includegraphics{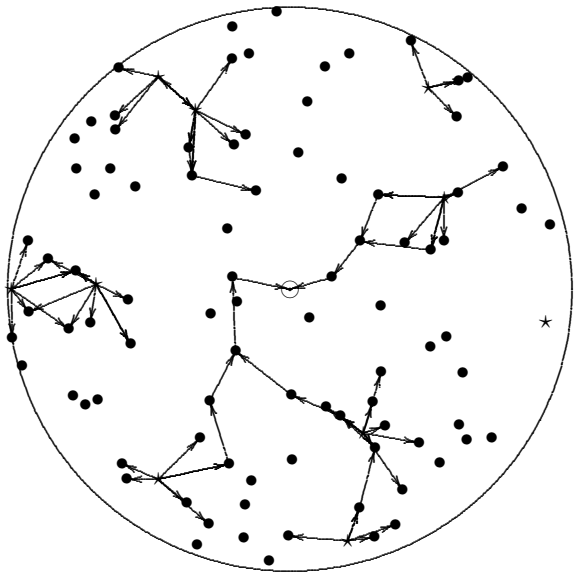}}
\caption{Digraph $D$ for a circle of radius $R$ centered on the sink with
$n=100$, $n^\star=10$, $f=0.3$, and $r\approx 0.26R$. The sink is represented by
$\odot$, each sensor by either $\bullet$ or, if a source, $\star$.}
\label{graph}
\end{figure}

Before moving on to simulation results, we pause for a pictorial illustration
of how the algorithm works. This illustration is shown in Figure~\ref{graph} for
an arrangement of sensors generated uniformly at random. What the figure shows
is a digraph whose set of nodes is the set of sensors enlarged by the sink, and
whose edges represent some of the messages exchanged during the execution of the
algorithm. Specifically, an edge exists either from sensor $i$ to the sink, if
an $\mathit{Answer}$ message is received by the sink from $i$ in Action~1; or
from sensor $j$ to sensor $i$, if $i$ broadcasts an $\mathit{Answer}$ message as
its timer goes off in Action~2 and $j$ is one of the sensors on which
information is present in $\mathcal{S}_i$ at the time of the broadcast. We
henceforth denote this digraph by $D$.

\section{Simulation results}\label{results}

In this section we report on selected results from extensive simulations of the
algorithm of Section~\ref{algo}. All our simulations were conducted on circles
of radius $R$ centered on the sink. For each simulation the $n$ sensors were
placed uniformly at random inside the circle and then the $n^\star$ sources were
selected also at random.

Every broadcast by a sensor is during a simulation assumed to reach exactly
those sensors that lie within a circle of radius $r$ centered on the emitting
sensor. The value of $r$ is determined so that the expected sensor density
inside the circle is the same as in the larger circle of radius $R$. If $n_r$
denotes the expected number of sensors inside the circle of radius $r$, then
we have $n_r/\pi r^2=n/\pi R^2$, so it follows that
\begin{equation}
r=\sqrt{\frac{n_r}{n}}R.
\label{rfromn}
\end{equation}
The parameter $r$ is then a function of $n_r$, so in our experiments the two
parameters that we vary are $f$ and $n_r$.

We evaluate the results of each simulation by means of the following three
indicators:
\begin{description}
\item[Fraction of connected sources.] The fraction, relative to $n^\star$,
representing the number of sources from which a directed path exists to the sink
in $D$. This indicator is a number in the interval $[0,1]$.
\item[Power usage ratio.] Since every sensor broadcasts with the same power, the
number of message broadcasts in Action~2 is proportional to the overall energy
expenditure by the sensors. The minimum number of broadcasts is $n^\star$ (one
for each source upon receiving the sink's $\mathit{Question}$ message), so this
indicator gives the ratio of the total number of broadcasts by sensors to
$n^\star$. This indicator is a number no less than $1$.
\item[Treeness.] Let $c$ be the number of nodes from which a directed path
exists in $D$ to the sink.  These nodes include the sink itself and are part of
the weakly connected component of $D$ that contains the sink.\footnote{A weakly
connected component of a digraph is any sub-digraph whose underlying undirected
graph is one of the connected components of the underlying undirected graph of
the digraph.} Clearly, the least possible number of edges lying on directed
paths from such nodes to the sink is $c-1$. This indicator gives the ratio of
the actual number of edges lying on directed paths to the sink to $c-1$. It is a
number no less than $1$ (though we assume it is $0$ when $c=1$, hence an average
may fall below $1$).
\end{description}

We show results for $n=2000$, $n^\star/n=a\times 10^{-b}$ with $a=1,2,5$ and
$b=1,2,3$, $f=0.1,0.3,0.5$, and $n_r=9,11,13,15$ (by (\ref{rfromn}), these
values of $n_r$ correspond to respectively
$r\approx 0.067R,0.074R,0.081R,0.087R$). For each combination of these values,
we give each of the three indicators as the average over $200$ independent
simulations. The results appear in the plots of Figure~\ref{plots}, which are
arranged into sets occupying four columns and three rows. Each of columns
(a)--(d) corresponds to a different value of $n_r$, which increases as we move
from (a) through (d). Each of the rows is specific to one of the three
indicators.

\begin{figure}
\centering
\begin{tabular}{c@{\hspace{0.20in}}c}
\scalebox{0.325}{\includegraphics{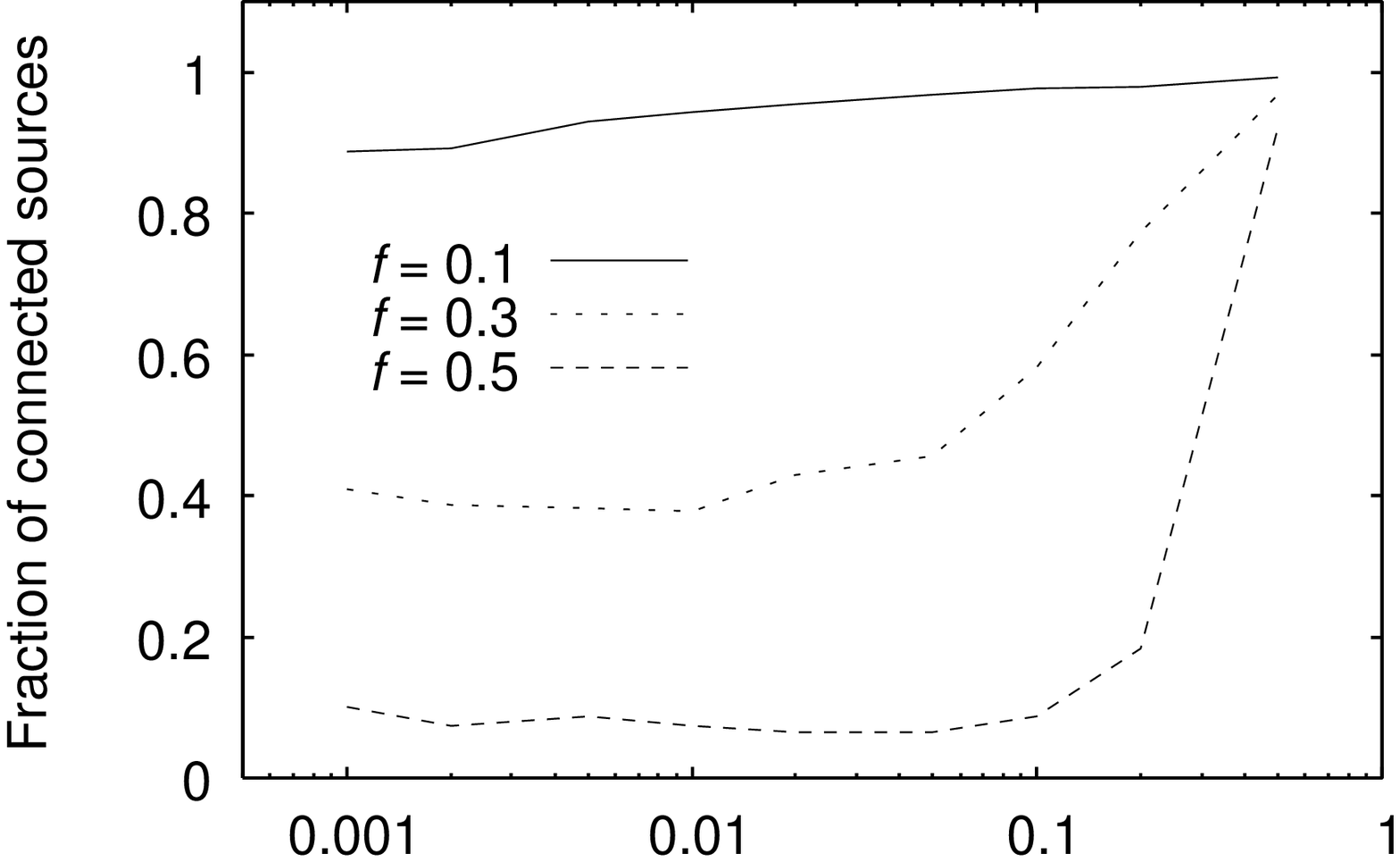}}&
\scalebox{0.325}{\includegraphics{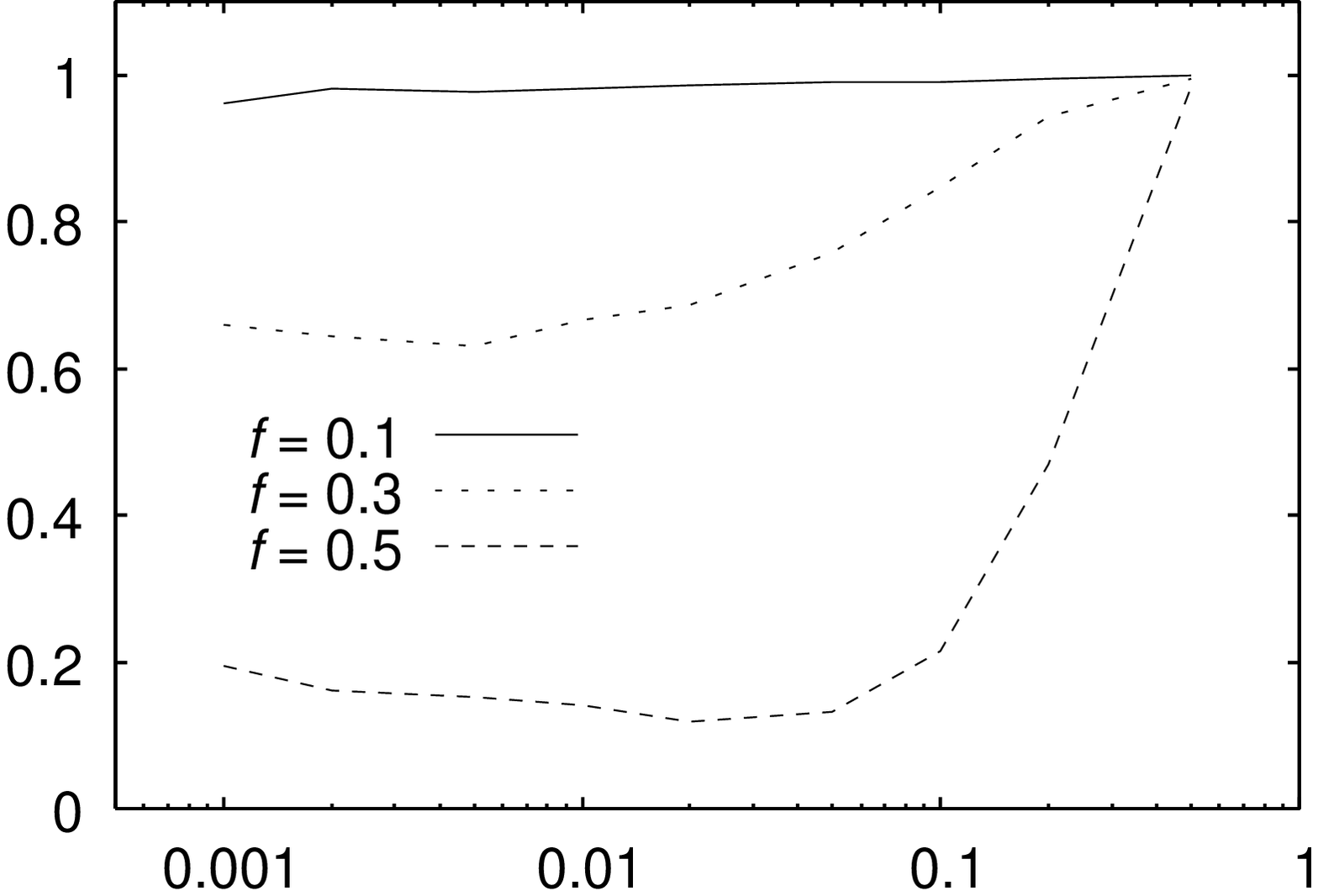}}\\
\scalebox{0.325}{\includegraphics{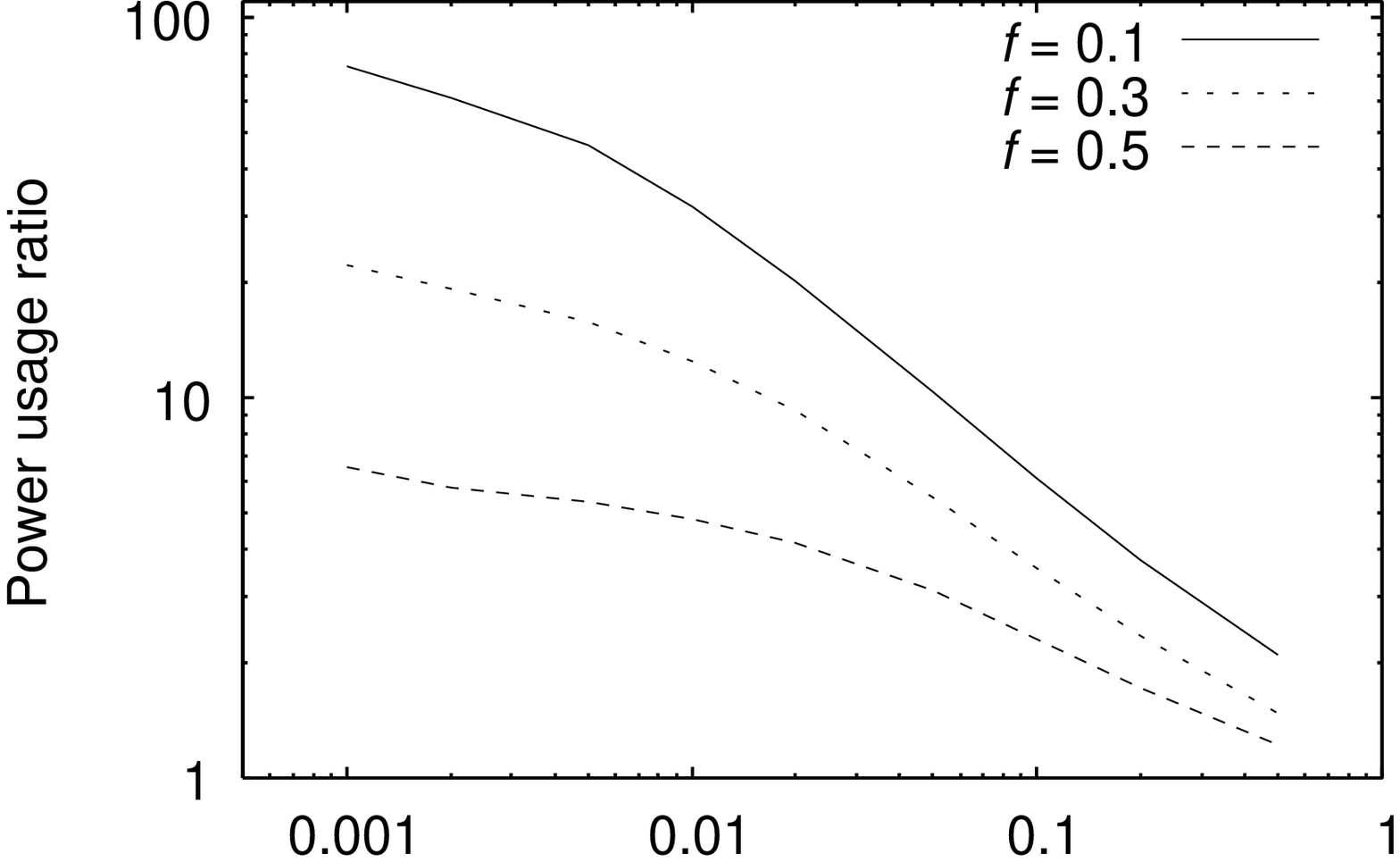}}&
\scalebox{0.325}{\includegraphics{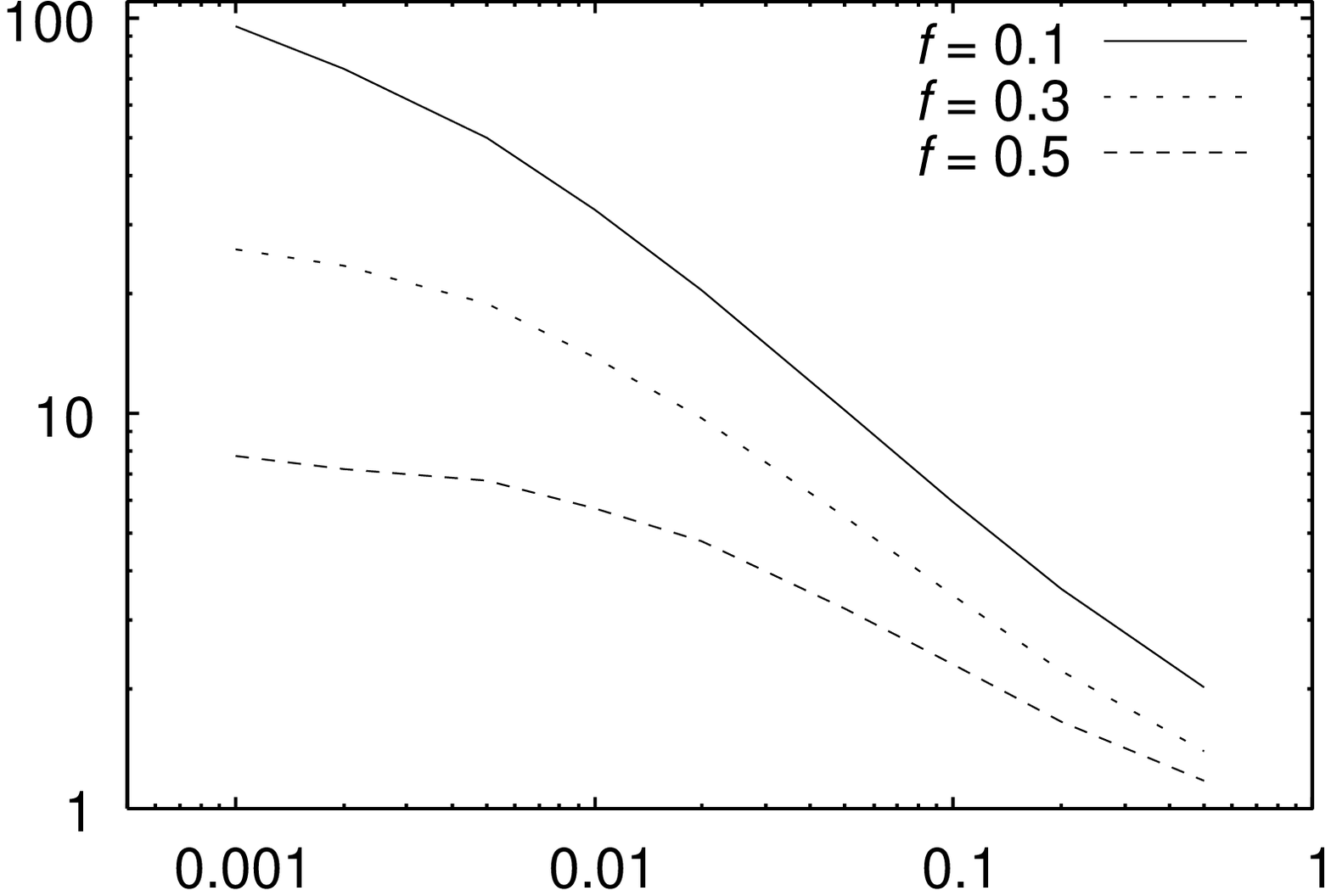}}\\
\scalebox{0.325}{\includegraphics{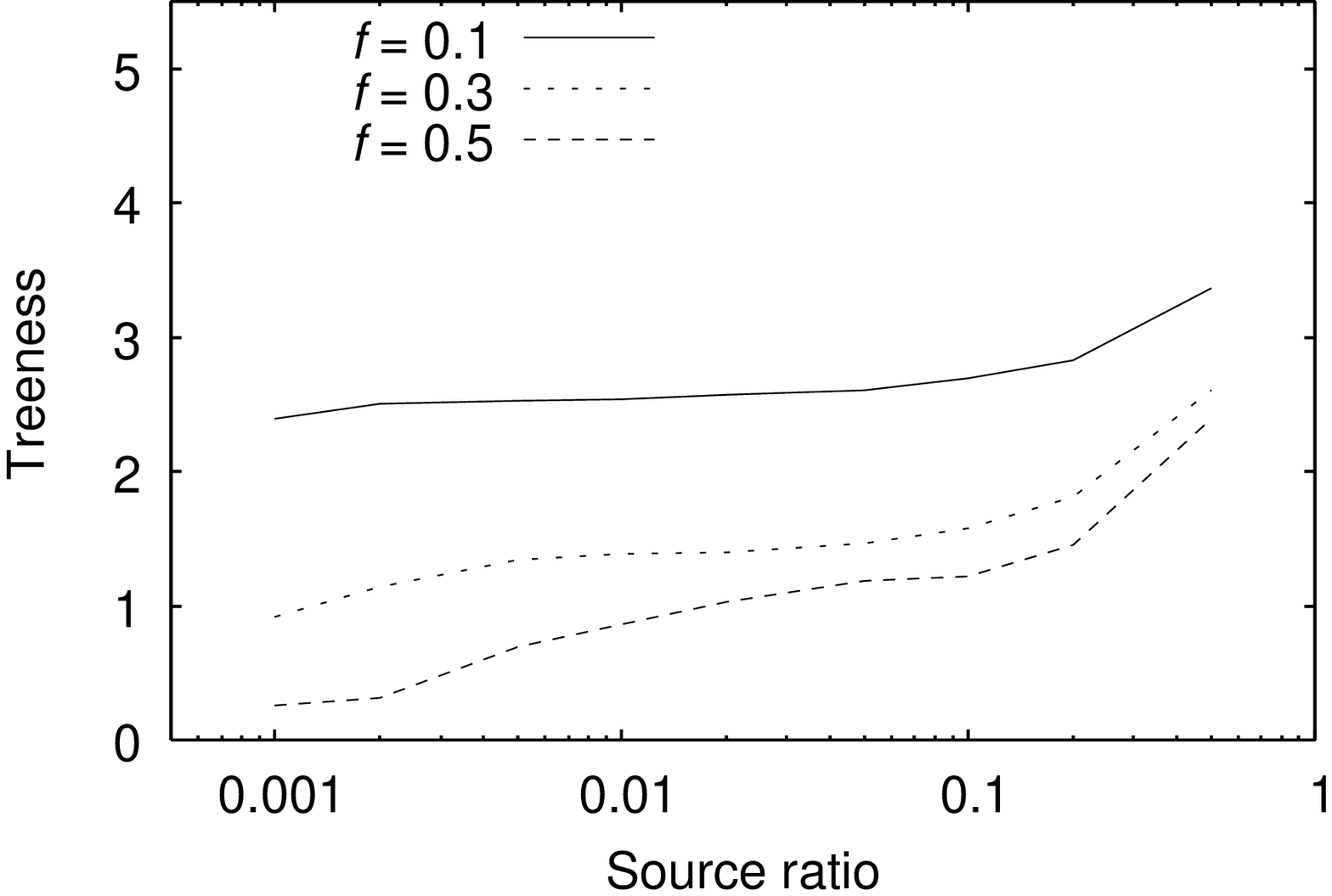}}&
\scalebox{0.325}{\includegraphics{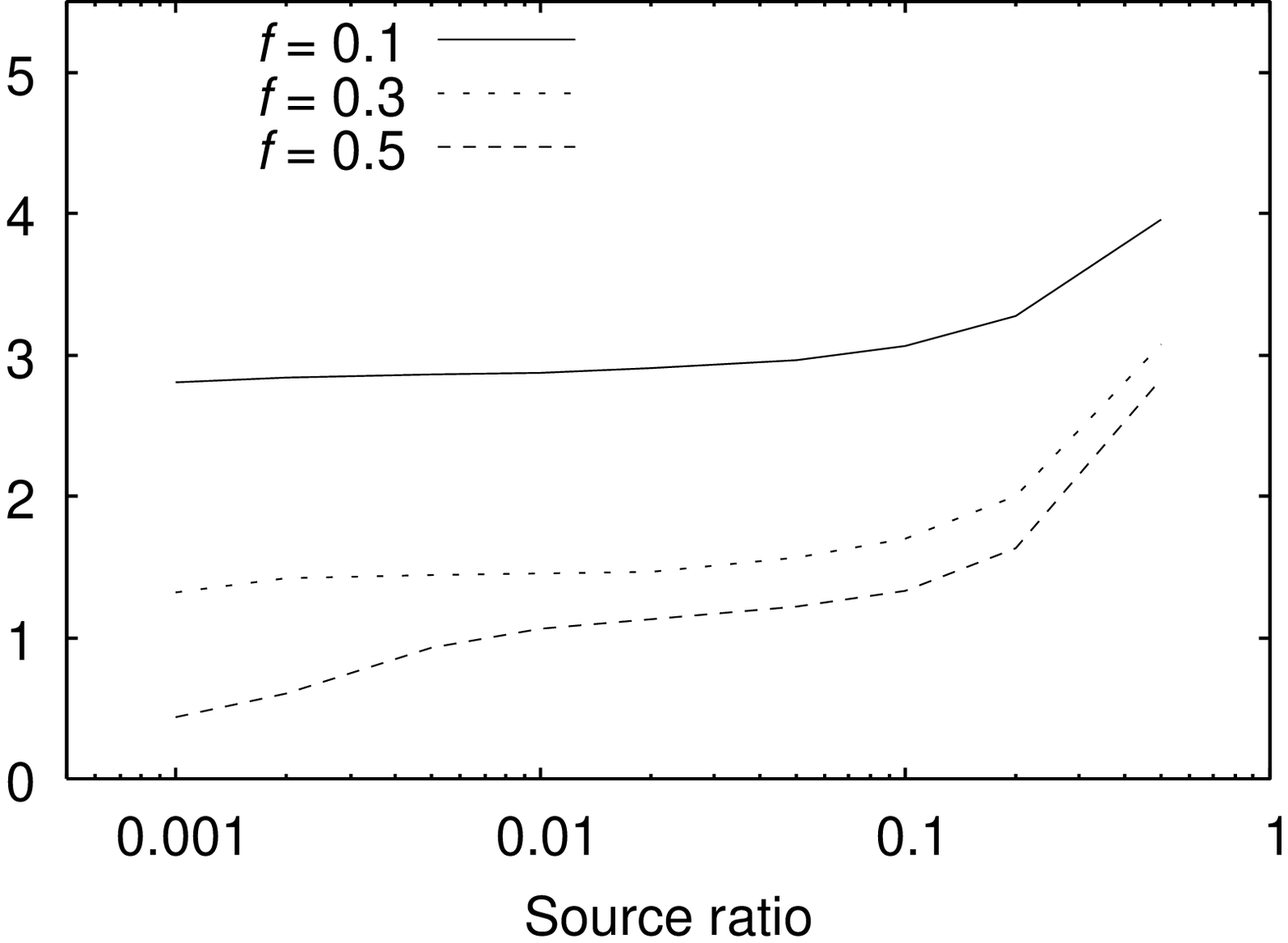}}\\
\footnotesize{(a)}&\footnotesize{(b)}
\end{tabular}
\caption{The three performance indicators plotted against the source ratio
$n^\star/n$: $n_r=9$ in part (a), $11$ in (b), $13$ in (c), and $15$ in (d).}
\label{plots}
\end{figure}

\addtocounter{figure}{-1}
\begin{figure}
\centering
\begin{tabular}{c@{\hspace{0.20in}}c}
\scalebox{0.325}{\includegraphics{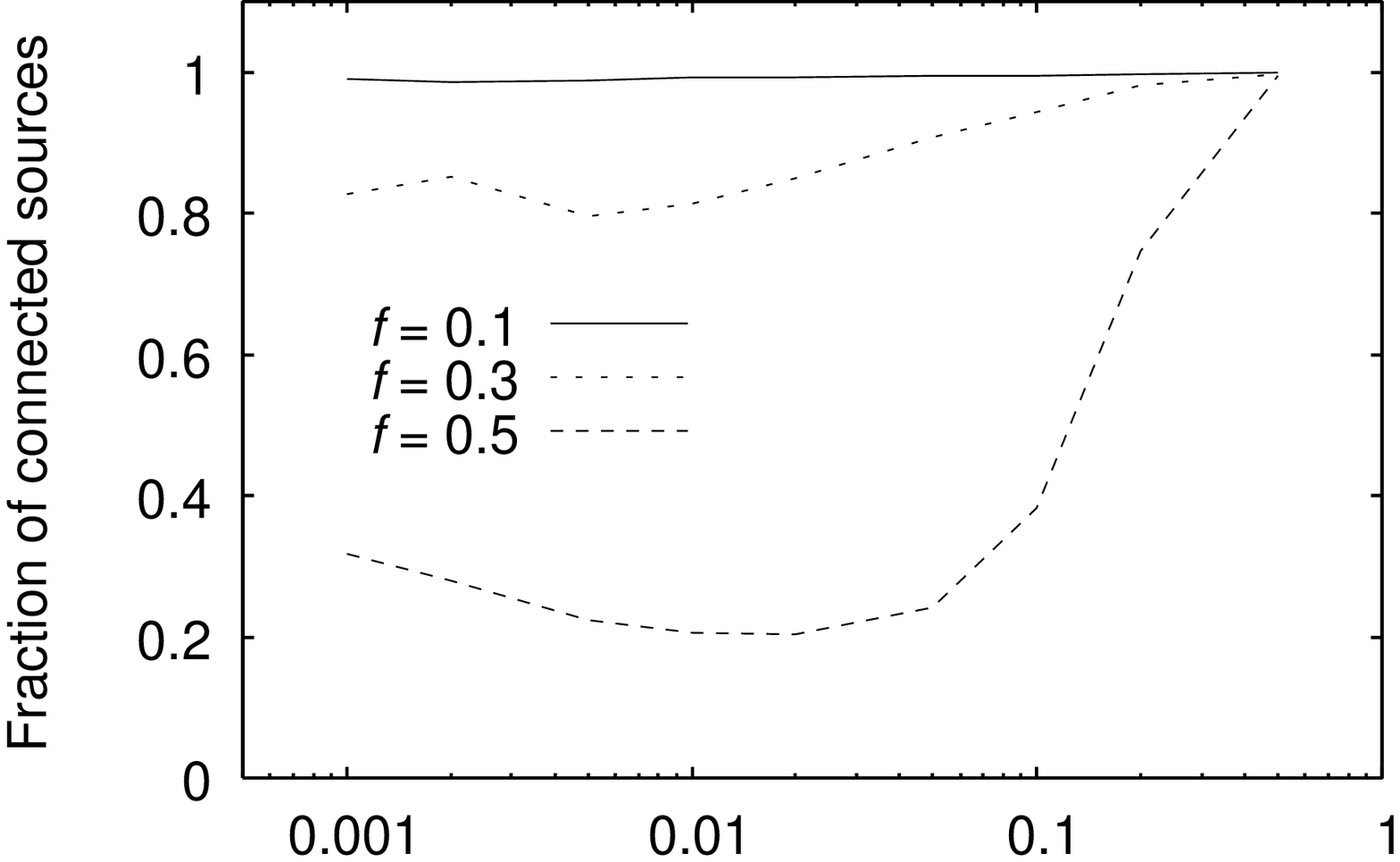}}&
\scalebox{0.325}{\includegraphics{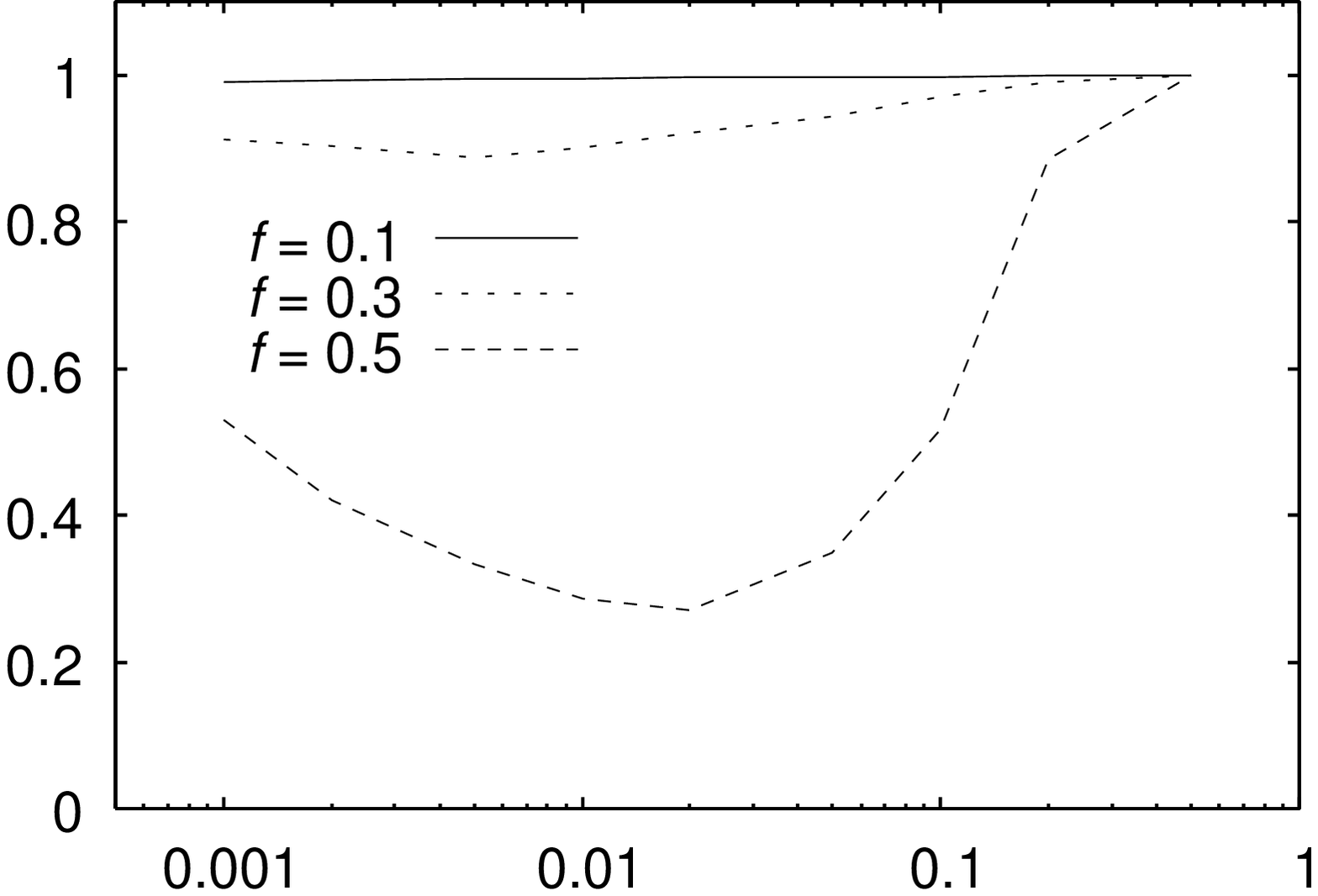}}\\
\scalebox{0.325}{\includegraphics{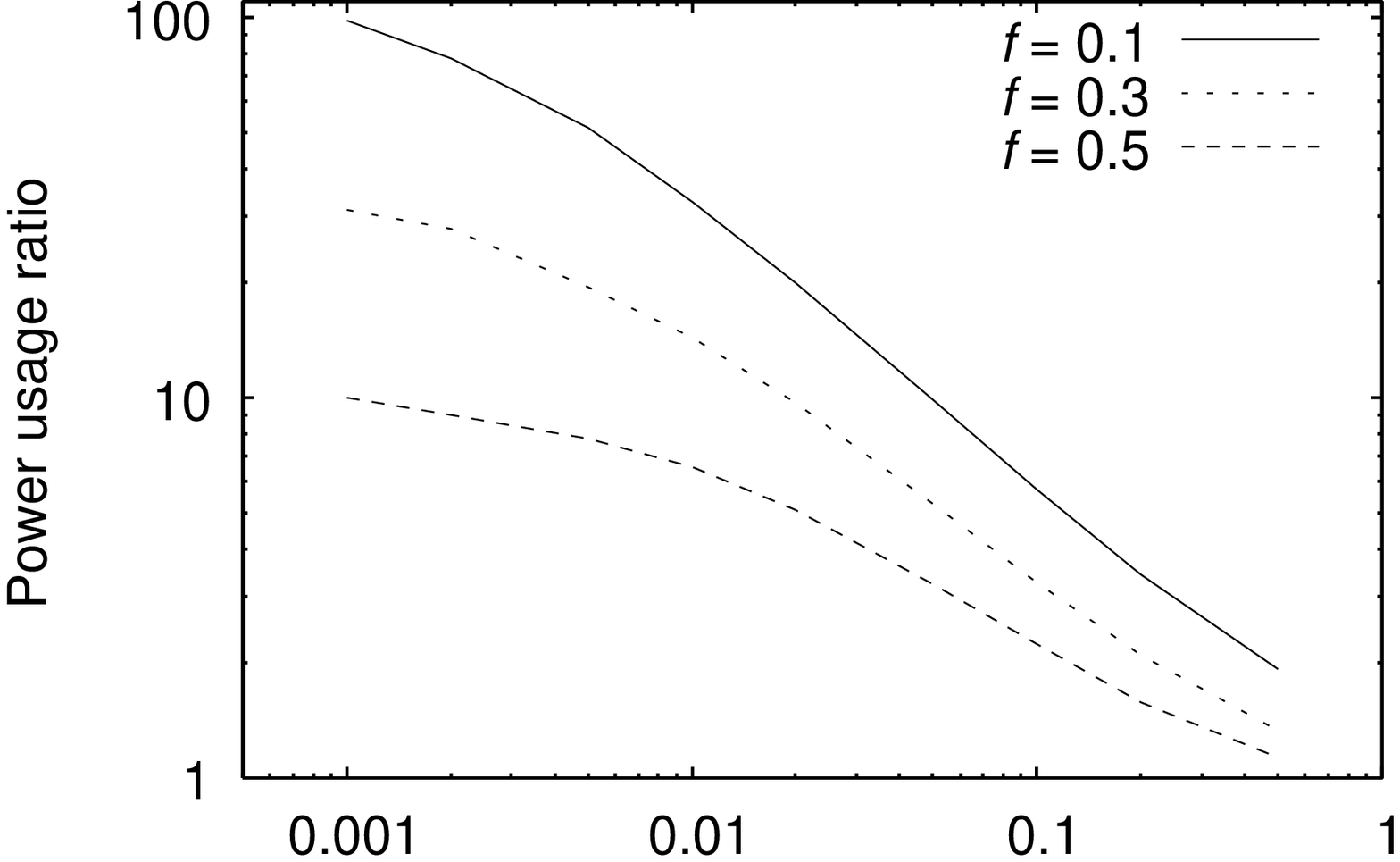}}&
\scalebox{0.325}{\includegraphics{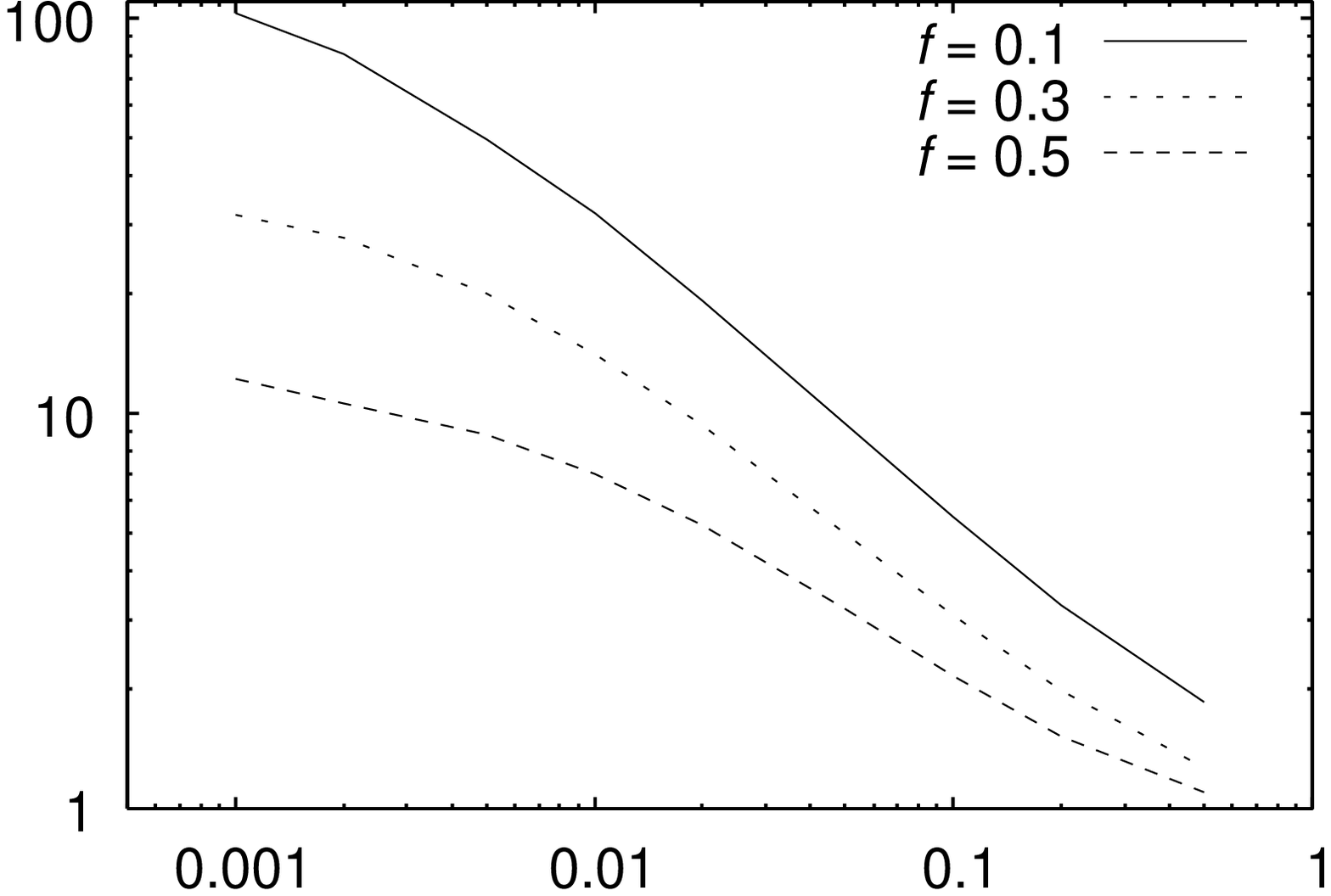}}\\
\scalebox{0.325}{\includegraphics{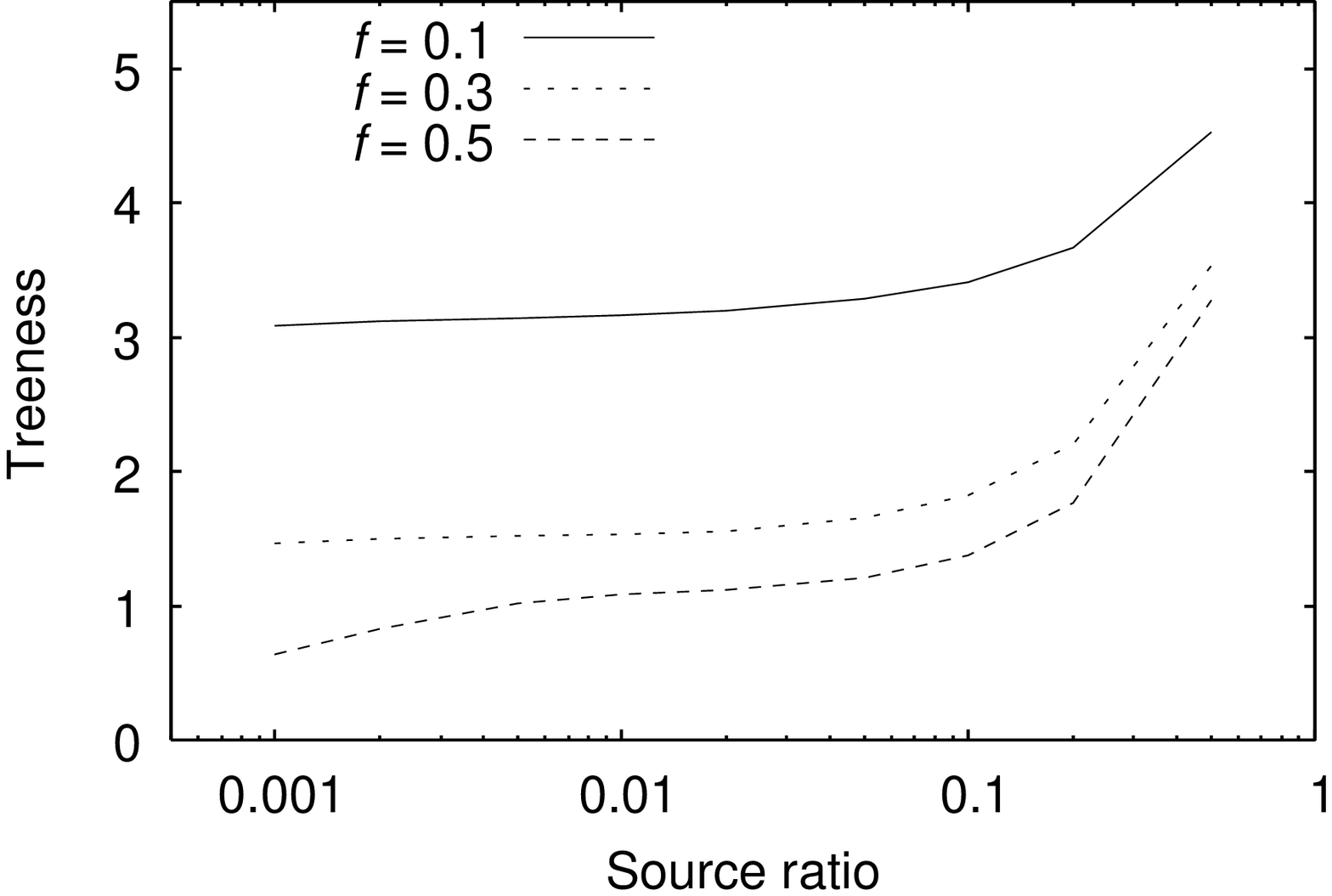}}&
\scalebox{0.325}{\includegraphics{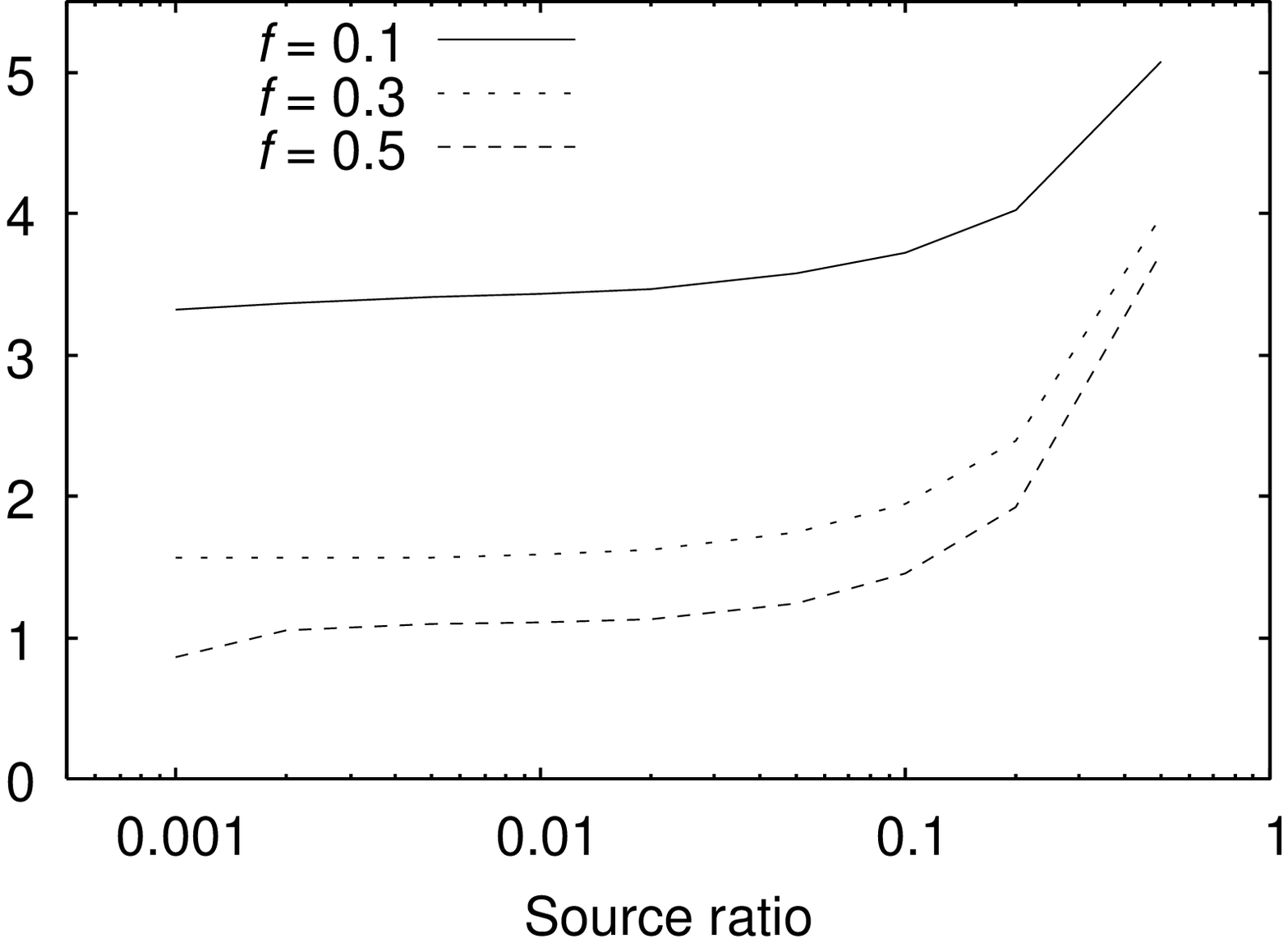}}\\
\footnotesize{(c)}&\footnotesize{(d)}
\end{tabular}
\caption{Continued.}
\end{figure}

As we move from (a) through (d) within the top row of plot sets, clearly the
fraction of connected sources improves as $n_r$ is increased. However, except
for the very dense case of $n^\star/n=0.5$, in which this indicator is
consistently very near $1$ regardless of the value of $f$, only values of $f$ no
larger than $0.3$ seem to yield acceptable performance. The combination of
$f=0.1$ and $n_r=13$, in particular, seems to already sustain a value of $1$
throughout the entire range of $n^\star/n$.

The middle row of plot sets indicates that increasing $n_r$ within the four
possibilities we have shown increases the power usage ratio only moderately (in
fact, close to negligibly for $f=0.1$, particularly near the upper end of $n_r$
values). One must bear in mind, however, that this indicator is only a relative
measure. The real energy expenditure involved grows with $r^2$ \cite{lc70},
therefore linearly with $n_r$, by (\ref{rfromn}). Other than this, the fact that
the four plot sets follow roughly the same functional forms is really easily
interpretable: at the lower end of $n^\star/n$ values, very few sensors are
sources, so conveying their answers to the sink must enlist the participation of
several other sensors for routing and aggregation and consequently the number of
broadcasts by sensors is many times larger than the number of broadcasts by
sources; at the upper end, half the sensors are sources, so broadcasts still
tend to occur in excess of broadcasts by sources but only moderately so.

The bottom row of plot sets refers to the treeness indicator and is therefore
related to assessing how many hops are needed in excess of a tree for conveying
to the sink the answers from the sources that really make it (i.e., those inside
$D$'s weakly connected component that includes the sink). All plots are roughly
flat within the middle interval of $n^\star/n$ values, particularly so for
smaller $f$ values, thus indicating that inside that interval increasing
$n^\star/n$ causes the component of $D$ to acquire more edges in approximately
the same rate as it acquires nodes. That treeness should be higher for lower
$f$, finally, is really to be expected, since lower $f$ means shorter (therefore
more redundant) hops.

\section{Conclusions}\label{concl}

We have considered the heretofore untouched question of building routes in
networks of anonymous sensors. We started with the basic premise that sensors
can measure how much power reaches them from the sink, and proposed a simple
distributed algorithm for building routes from sources to the sink that uses
such measurements as a means of providing some differentiation among the
sensors. The algorithm assumes an idealized broadcast model for the sink and the
sensors, but adapting it to a more realistic setting is expected to be a
relatively simple task.

We have provided simulation results that, in our understanding, are both
surprising and encouraging. In particular, they seem to suggest that the
radially decaying power perceived by the sensors as we move farther away from
the emitting sink is capable of sustaining the construction of routes from
randomly placed sources back to the sink. This is all achieved in the absence
of unique sensor identifications, so what we really observe is that our simple
algorithm leads the sensors to self-organize into conveying useful information
to the sink.

\subsection*{Acknowledgments}

The authors acknowledge partial support from CNPq, CAPES, and a FAPERJ BBP
grant.

\bibliography{sensornet}
\bibliographystyle{plain}

\end{document}